\documentclass[preprint,showpacs,showkeys,amssymb,superscriptaddress,nofootinbib]{revtex4}

\usepackage{amsmath}
\usepackage{graphicx}
\usepackage{epsfig}
\usepackage{dcolumn}
\usepackage{bm}

\usepackage{epsfig}
\def\gsim{\lower0.5ex\hbox{$\:\buildrel >\over\sim\:$}}
\def\lsim{\lower0.5ex\hbox{$\:\buildrel <\over\sim\:$}}
\newcommand{\bea}{\begin{eqnarray}}
\newcommand{\eea}{\end{eqnarray}}

\begin{document}

%
%
\begin{flushright}
\hspace*{5.5in}
\mbox{HIP-2011-16/TH}
\end{flushright}
\title{Sneutrino-antisneutrino oscillation at the Tevatron}
%
\author{Dilip Kumar Ghosh}%
\email{tpdkg@iacs.res.in}
\affiliation{Department of Theoretical Physics, 
Indian Association for the Cultivation of Science, 2A $\&$ 2B Raja
S.C. Mullick Road, Kolkata 700 032, India}
\author{Tuomas Honkavaara}%
\email{Tuomas.Honkavaara@helsinki.fi}
\affiliation{Department of Physics, and Helsinki Institute of Physics,
P.O. Box 64, FIN-00014 University of Helsinki, Finland}
\author{Katri Huitu}%
\email{Katri.Huitu@helsinki.fi}
\affiliation{Department of Physics, and Helsinki Institute of Physics, P.O. Box 64, FIN-00014 University of Helsinki, Finland}
\author{Sourov Roy}%
\email{tpsr@iacs.res.in}
\affiliation{Department of Theoretical Physics, Indian Association for the Cultivation of Science, 2A $\&$ 2B Raja S.C. Mullick Road, Kolkata 700 032, India}
\date{\today}
\begin{abstract}
Sneutrino-antisneutrino oscillation can be a very useful probe to look for signatures of lepton number violation ($\Delta L=2$) at the Tevatron. Here, we discuss a scenario where the total decay width of the sneutrino is very small, producing interesting signals at the Tevatron for a mass splitting $\Delta m$ as small as $10^{-14}$ GeV between the sneutrino mass eigenstates.
\end{abstract}

\pacs{12.60.Jv, 14.60.Pq, 14.80.Ly}

\keywords{Supersymmetry, sneutrino, sneutrino-antisneutrino oscillation}

\maketitle


From the neutrino oscillation experiments, we have compelling evidence that neutrinos have tiny, nonzero masses. Whenever the neutrinos are of the Majorana type, sneutrino-antisneutrino mixing occurs in any supersymmetric model. Such $\Delta L=2$ Majorana neutrino mass terms can induce a mass splitting ($\Delta m_{\tilde{\nu}}$) between the physical states. This mass splitting induces sneutrino-antisneutrino oscillations which probe the lepton number violation ($\Delta L=2$) \cite{hirschetal,grossman-haber,choietal,chun,tuomas-egamma,dedes-haber-rosiek,tuomas-lhc}. These oscillations are analogous to $B^0$--$\bar{B}^0$ and $K^0$--$\bar{K}^0$ oscillations. Here, we assume that the sneutrino flavour oscillation is absent and lepton flavour is conserved in the decay of the sneutrino/antisneutrino.

Usually, the oscillation probability formula where the sneutrinos are produced at rest is used. However, it was discussed in \cite{new-formula-lhc} that, for a very small sneutrino decay width $\Gamma$, say, $\sim 10^{-14}$ GeV, it is very important to take into account the effects of the Lorentz factor $\gamma$ and the distance $L$ (at which the measurement is made inside the detector) in the calculation 
of the sneutrino-antisneutrino oscillation probability. 
Since the sneutrinos (antisneutrinos) decay, we need to look at the integrated probability. Hence, the integrated probability, at a distance $L$, of a $|\tilde \nu\rangle$ oscillating into an $|{\tilde \nu}^*\rangle$ is given by \cite{new-formula-lhc}
\bea
P(L) &=& \frac{\int_0^L dx |\langle{\tilde \nu}^*|\psi(x,t)\rangle|^2}
{\int_0^\infty dx \langle \psi(x,t)|\psi(x,t)\rangle} \nonumber \\
&=& \frac{e^{-L\alpha}}{2 (\alpha^2 + \beta^2)}\Big[-\alpha^2 + (-1 + e^{L\alpha})\beta^2 
 +\alpha^2 \cos(L\beta) - \alpha \beta \sin(L\beta)\Big], 
\label{length_dependent_int_osc_formula}
\eea
where $\alpha \equiv \frac{\Gamma m}{E}$ and $\beta \equiv \frac{\Delta m^2}{2E}$. For a very large $L$, i.e., when $L \alpha \gg$ 1, we get from Eq.\ (\ref{length_dependent_int_osc_formula}) that
\bea
P(L) = \frac{\beta^2}{2 (\alpha^2 + \beta^2)} = 
\frac{x^2_{\tilde \nu}}{2(1 + x^2_{\tilde \nu})},
 \label{old_osc_prob}
\eea
which is independent of $L$ and where we use the relations $\Delta m^2 = 2m \Delta m$ and 
$x_{\tilde \nu} \equiv \frac{\Delta m}{\Gamma}$ \cite{grossman-haber}. 
Equation (\ref{old_osc_prob}) is the same result as in the case when the sneutrinos are 
produced at rest. Worth noting from Eq.\ (\ref{old_osc_prob}) is that, 
when $x_{\tilde \nu}$ = 1, the oscillation probability $P({\tilde \nu} \rightarrow {\tilde \nu}^*)$ is 0.25. On the other 
hand, when $x_{\tilde \nu} \gg$ 1, $P({\tilde \nu} \rightarrow {\tilde \nu}^*)$ is 0.5. The general effects of the Lorentz factor
$\gamma=\frac{E}{m}$, $L$, and $\Gamma$ on the oscillation probability have been studied in \cite{new-formula-lhc}. 

In this paper, we study sneutrino-antisneutrino oscillation at the Tevatron collider in a 
scenario where the sneutrino width is very small ($\sim 10^{-14}$ GeV). Even though the Lorentz factor $\gamma$ for the sneutrinos at the Tevatron is not as high as it can be at the LHC, the tiny width makes it important to use the general formula in Eq.\ (\ref{length_dependent_int_osc_formula}), since the $L$- and $\gamma$-dependences can be much more pronounced.
We illustrate this in the context of an interesting supersymmetric scenario which can produce spectacular signals at the Tevatron.

These small values of the sneutrino decay width of the order of $10^{-14}$ GeV are possible, for example, in a scenario where the left-handed sneutrino next-to-lightest supersymmetric particle is nearly degenerate to the lighter stau lightest supersymmetric particle and the dominant decay channel for ${\tilde \nu}_\tau$ is 
\bea
{\tilde \nu}_\tau \rightarrow {\tilde \tau}_1^- + \pi^+,
\eea
with a total decay width $\Gamma \sim 10^{-14}$ GeV. In some models with an extra $U(1)_{B-L}$, the oscillation of a right-chiral sneutrino (${\tilde \nu}_R$) can be important \cite{khalil_B-L_RHSNU_osc}. In such cases, the total decay width of ${\tilde \nu}_R$ can be again as small as $\sim 10^{-14}$ GeV. The left-chiral sneutrino decay width can also be reduced if it has a significant mixing with the right-chiral counterpart.

We study the case when the dominant sneutrino decay is ${\tilde \nu}_\tau \rightarrow {\tilde \tau}_1^- + \pi^+$. Then, at the Tevatron, a possible signal would be $p\bar{p} \rightarrow {\tilde \nu}_\tau {\tilde \tau}_1^+ \rightarrow {\tilde \tau}_1^- {\tilde \tau}_1^+ + \pi^+$. For the staus, we assume a small $R$-parity violating coupling ($\lsim 10^{-8}$) such that the $\tilde{\tau}_1$ decays outside the detector, leaving a heavily ionized charged track. We assume that this small $R$-parity violating coupling does not change the total decay width of the sneutrino. This means that the above signal produces two heavily ionized charged tracks with opposite curvatures when there is no sneutrino oscillation and with same curvatures when there is sneutrino oscillation. We assume that, due to 
slower velocity of staus, these stau tracks can be distinguished from the muon tracks. Similarly, one should also look at the signal $p\bar{p} \rightarrow {\tilde \nu}^*_\tau {\tilde \tau}_1^- \rightarrow {\tilde \tau}_1^+ {\tilde \tau}_1^- + \pi^-$. Worth noting here is that the sneutrino is long-lived (decay length a bit under a centimeter), and, hence, 
one of the staus produced from the decay of the sneutrino shows a secondary vertex which is well separated from the primary vertex. This is a very spectacular signal and free from any standard model or supersymmetric backgrounds. 

Now, we discuss the cross section and the branching ratio of the processes 
discussed above. We consider $\tilde{\nu}_\tau$ as the next-to-lightest supersymmetric particle and $\tilde{\tau}_1$ as 
the lightest supersymmetric particle. As mentioned earlier, due to the tiny $R$-parity violating coupling, the $\tilde{\tau}_1$ decays outside the detector, leaving a heavily ionized charged track. The mass of the sneutrino is considered to be $m_{{\tilde \nu}_\tau} = 100$ GeV and the mass of ${\tilde \tau}_1$ is $m_{{\tilde \tau}_1} = 99.7$ GeV. The stau mixing angle is taken to be $\pi/4$. The other relevant 
parameter choices are $M_1 = 120$ GeV, $M_2 = 240$ GeV, $\mu = -392$ GeV, $\tan\beta=8$, $m_{A^0}=600$ GeV and $A_\tau = 390$ GeV\footnote{There were mistakes in some of these numbers in \cite{new-formula-lhc}. However, the results remain unchanged.}. Here, $M_1$ and $M_2$ are the $U(1)$ and $SU(2)$ gaugino mass parameters, respectively, $\mu$ is the superpotential $\mu$-parameter, $m_{A^0}$ is the pseudoscalar Higgs boson mass and $A_\tau$ is the trilinear scalar coupling of the staus. With these values of parameters, the total decay width of the sneutrino is $\Gamma \approx 1 \times 10^{-14}$ GeV, while
the branching ratio of the decay ${\tilde \nu}_\tau \rightarrow {\tilde \tau}_1^- + \pi^+$ is 93$\%$. In fact, the branching ratio is greater than 90$\%$ when the mass splitting between the ${\tilde \nu}_\tau$ and the ${\tilde \tau}_1$ is in the range 200--350 MeV. 
We get the opposite-sign stau signal (OS) $p\bar{p} \rightarrow {\tilde \tau}_1^+ {\tilde \tau}_1^-$ from both ${\tilde \nu}_\tau {\tilde \tau}_1^+$ and ${\tilde \nu}^*_\tau {\tilde \tau}_1^-$ productions with an {\it effective} survival probability $(1-P_\mathrm{eff})$. The same-sign stau signal (SS) $p\bar{p} \rightarrow {\tilde \tau}_1^+ {\tilde \tau}_1^+$ or ${\tilde \tau}_1^- {\tilde \tau}_1^-$ we get from either ${\tilde \nu}_\tau {\tilde \tau}_1^+$ or ${\tilde \nu}^*_\tau {\tilde \tau}_1^-$ productions with the {\it effective} oscillation probability $(P_\mathrm{eff})$.

We select the signal events with the following criteria: 
1) the pseudorapidities of the staus 
must be $|\eta^{{\tilde \tau}_1}|< 1.2$ \cite{cuts}, 
2) the transverse momentum of both staus must satisfy 
$p^{{\tilde \tau}_1}_T> 20$ GeV, 
3) the isolation variable $\Delta R \equiv \sqrt{(\Delta \eta)^2 + (\Delta \phi)^2}$ should satisfy $\Delta R> 1.0$ for the two staus, and
4) the $\beta\gamma$ should be $0.3 < \beta\gamma < 2.0$, as noticed for the LHC in \cite {new-formula-lhc}. 
The upper limit of $\beta\gamma$ reduces the muon background considerably. Applying these cuts, the cross sections for the parameter point mentioned earlier with different $\Delta m$'s for $L=0.08$ m and $L=0.28$ m are presented in Table \ref{tab_cross_sec}.
\begin{table}
\begin{center}
\footnotesize
\begin{tabular}{|c|c|c|c|c|c|c|}
\hline
$\Delta m$ [GeV] & \multicolumn{2}{c|}{$10^{-14}$} & \multicolumn{2}{c|}{$10^{-13}$} & \multicolumn{2}{c|}{$10^{-10}$} \\
\hline
 & \multicolumn{6}{c|}{Cross section in fb} \\
\hline
Signal & OS & SS & OS & SS & OS & SS \\
\hline
$L=0.08$ m & 11.0 & 2.4 & 7.3 & 6.2 & 7.2 & 6.3 \\
\hline
$L=0.28$ m & 10.3 & 3.2 & 6.8 & 6.6 & 6.7 & 6.7 \\
\hline
\end{tabular}
\end{center}
\caption{Cross sections for the OS and SS stau signals with different $L$'s and $\Delta m$'s for the parameter point discussed in the text. The cuts used are also mentioned in the text.}
\label{tab_cross_sec}
\end{table}
It can be seen from this table that, for $\Delta m \gtrsim 10^{-13}$ GeV, 
the cross sections almost saturate. Even putting $\Delta m$ to its 
maximum value, $10^{-7}$ GeV (this comes from Eq.\ (8) of 
Ref.\ \cite{grossman-haber} assuming that 
$m_{\nu_\tau}=\mathcal{O}(0.1)$ eV), does not change the 
cross sections from $\Delta m = 10^{-10}$ GeV values. 
On the other hand, we can probe down to $\Delta m = 10^{-14}$ GeV 
and measure several SS events with 10 fb$^{-1}$ luminosity. 
Using the SS and OS cross sections, we define the asymmetry 
$ A = \frac{\sigma({\rm SS}) - \sigma({\rm OS})}{\sigma({\rm SS}) + \sigma({\rm OS})}$. 
For illustration, we show one value of this asymmetry, $A = -0.067\pm 0.086$, 
obtained by using $\sigma_{\rm SS}$ and $\sigma_{\rm OS}$ for 
$L=0.08$ m with $\Delta m = 10^{-10}$ GeV from 
Table \ref{tab_cross_sec} and assuming an integrated luminosity of 
$10~{\rm fb}^{-1}$. The asymmetry is almost consistent with zero, simply because of lower statistics
available at the given luminosity at Tevatron energy (corresponds to larger error) and larger oscillation probability 
(smaller central value of the asymmetry).
The asymmetry $A$ gives direct information about the oscillation probability and is independent of initial state parton densities and other uncertainities arising from higher order corrections. Easily, it can be checked that $P_{\rm eff} = (1+A)/2$. By measuring the value of $A$, one can calculate the {\it effective} oscillation probability. 
For our example, we get $P_{\rm eff} = 0.47$.

\begin{figure}
\begin{center}
\includegraphics[scale=0.35]{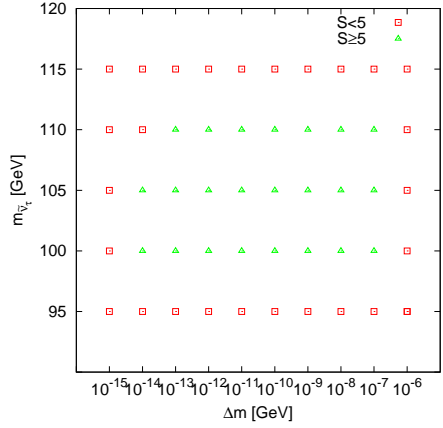}
\caption{Exclusion plot for $L=0.28$ m, $m_{\tilde{\nu}_\tau}-m_{\tilde{\tau}_1}=0.2$ GeV, and $\mathcal{L}=6$ fb$^{-1}$, showing in $(\Delta m,m_{\tilde{\nu}_\tau})$ plane whether the signal significance $S$ is greater or smaller than 5.} 
\label{exclusion_1}
\end{center}
\end{figure}

\begin{figure}
\begin{center}
\includegraphics[scale=0.35]{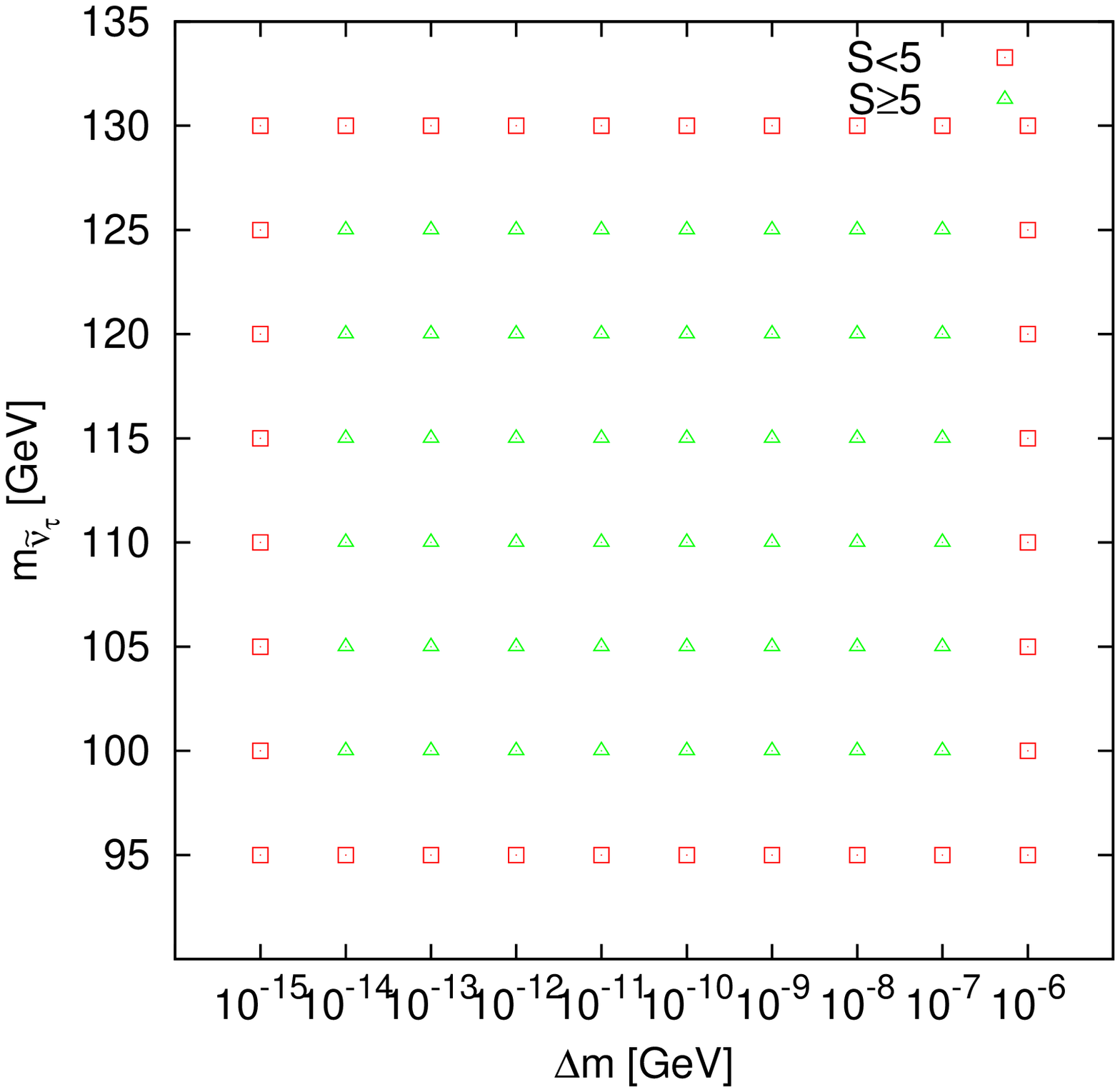}
\caption{Exclusion plot for $L=0.28$ m, $m_{\tilde{\nu}_\tau}-m_{\tilde{\tau}_1}=0.2$ GeV, and $\mathcal{L}=10$ fb$^{-1}$, showing in $(\Delta m,m_{\tilde{\nu}_\tau})$ plane whether the signal significance $S$ is greater or smaller than 5.} 
\label{exclusion_2}
\end{center}
\end{figure}

\begin{figure}
\begin{center}
\includegraphics[scale=0.35]{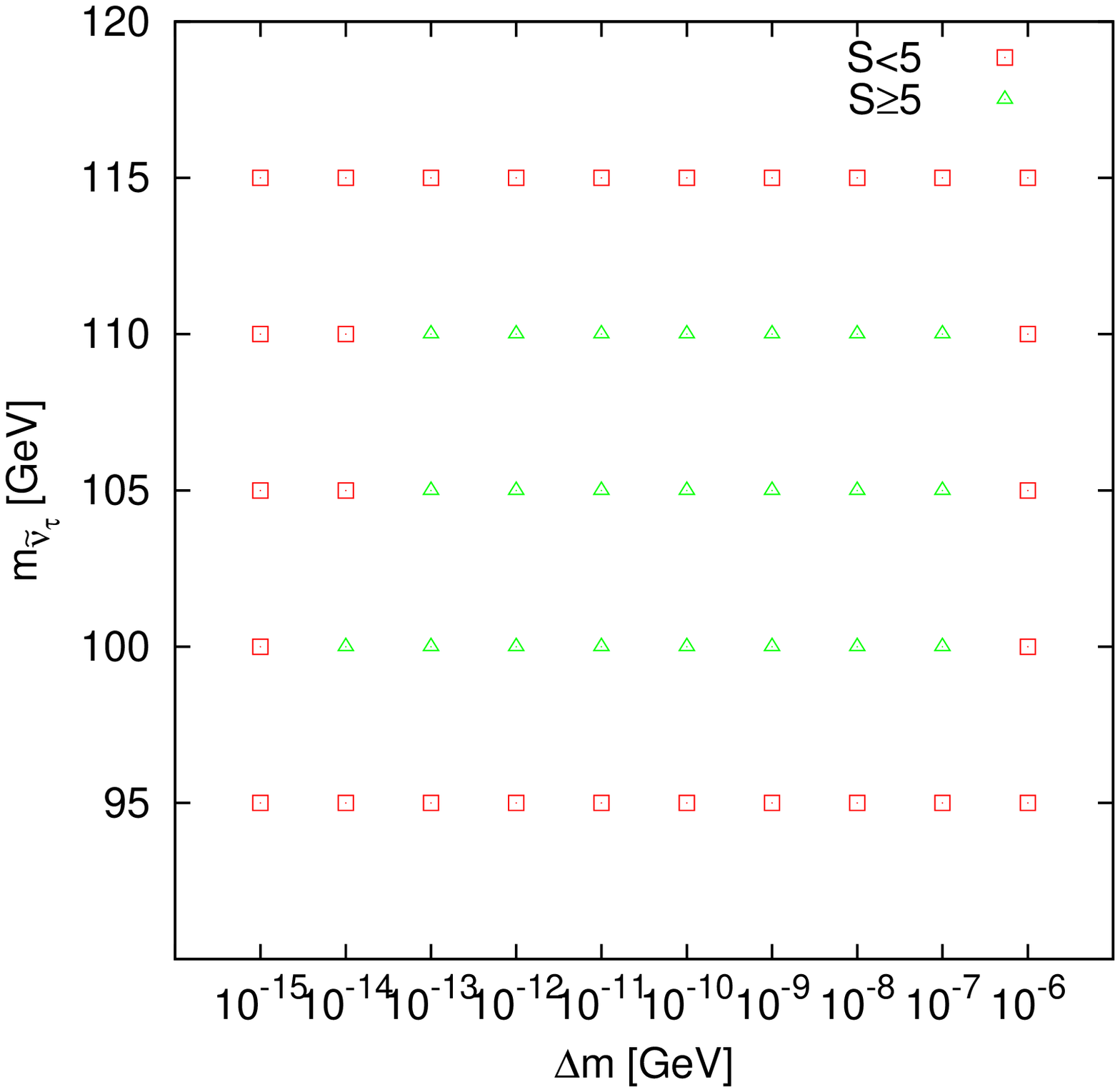}
\caption{Exclusion plot for $L=0.08$ m, $m_{\tilde{\nu}_\tau}-m_{\tilde{\tau}_1}=0.2$ GeV, and $\mathcal{L}=10$ fb$^{-1}$, showing in $(\Delta m,m_{\tilde{\nu}_\tau})$ plane whether the signal significance $S$ is greater or smaller than 5.} 
\label{exclusion_4}
\end{center}
\end{figure}

In Figs.\ \ref{exclusion_1}--\ref{exclusion_4}, there are exclusion plots, i.e., plots showing whether we can have the signal significance 
$S=N_S/\sqrt{N_S + N_B} \approx N_S/\sqrt{N_S} \geq 5$ for different $\Delta m$'s and $m_{\tilde{\nu}_\tau}$'s for different values of 
$m_{\tilde{\nu}_\tau}-m_{\tilde{\tau}_1}$. Here, $N_S$ is the number of signal events coming from SS. In all the cases, 
stau mixing is maximal, i.e., the mixing angle is $\pi/4$. 
In Figs.\ \ref{exclusion_1} and \ref{exclusion_2}, we show 
$L=0.28$ m, $m_{\tilde{\nu}_\tau}-m_{\tilde{\tau}_1}=0.2$ GeV,
and luminosity $\mathcal{L}$ is 6 fb$^{-1}$ and 10 fb$^{-1}$, 
respectively. It can be seen that, naturally, with higher 
luminosity, we can reach higher tau-sneutrino masses. Remember 
that, as mentioned earlier, the maximum value of $\Delta m$ is 
$10^{-7}$ GeV. Then, in Fig.\ \ref{exclusion_4}, we have changed 
$L$ to 0.08 m and, now, $\mathcal{L}=10$ fb$^{-1}$. It can be seen 
that we reach much lighter $m_{\tilde{\nu}_\tau}$'s. For 6 fb$^{-1}$, none of the points is allowed. The reason for this is that $L\alpha$ in Eq.\ \eqref{length_dependent_int_osc_formula} 
is somewhat larger than 1 for $L=0.28$ m. For the $L=0.08$ m case, 
$L\alpha<1$ for most of the cases, hence showing the effects of the general oscillation probability formula of Eq.\ \eqref{length_dependent_int_osc_formula}. 

When we change $m_{\tilde{\nu}_\tau}-m_{\tilde{\tau}_1}$ to 0.3 GeV, the width of the pion mode (${\tilde \nu}_\tau \rightarrow {\tilde \tau}_1^- + \pi^+$) grows. However, the width for the decays $\tilde{\nu}_\tau \to \tilde{\tau}_1^- \ell^+ \nu_\ell$, where $\ell=e,\mu$, grows even more, making the pion mode branching ratio drop a bit. 
The results for the $m_{\tilde{\nu}_\tau}-m_{\tilde{\tau}_1}=0.3$ GeV are rather similar to the $L=0.28$ m, $m_{\tilde{\nu}_\tau}-m_{\tilde{\tau}_1}=0.2$ GeV, and $\mathcal{L}=6 \ \mathrm{or} \ 10$ fb$^{-1}$ case. The descriptions for other exlusion regions are given in Table \ref{other_plots}.
When we change to $m_{\tilde{\nu}_\tau}-m_{\tilde{\tau}_1}=0.4$ GeV, the widths still grow, but the pion mode branching ratio drops a bit.  
\begin{table*}[!ht]
\begin{center}
\footnotesize
\begin{tabular}{|c|c|c|c|}
\hline
$L$ [m] & $m_{\tilde{\nu}_\tau}-m_{\tilde{\tau}_1}$ [GeV] & $\mathcal{L}$ [fb$^{-1}$] & Comment \\
\hline
0.28 & 0.3 & 6 & Figure \ref{exclusion_1}, except $\Delta m=10^{-14}$ GeV values disallowed \\
\hline
0.28 & 0.3 & 10 & Figure \ref{exclusion_2}, $\Delta m=10^{-14}$ GeV disallowed, except $m_{\tilde{\nu}_\tau}=100, 105$ GeV \\
\hline
0.08 & 0.3 & 6 & Figure \ref{exclusion_1}, except $\Delta m=10^{-14}$ GeV values disallowed \\
\hline
0.08 & 0.3 & 10 & Figure \ref{exclusion_2}, $\Delta m=10^{-14}$ GeV values disallowed \\
\hline
0.28 & 0.4 & 6 & Figure \ref{exclusion_1}, except $\Delta m=10^{-14}$ GeV values disallowed \\
\hline
0.28 & 0.4 & 10 & Figure \ref{exclusion_2}, $\Delta m=10^{-14}$ GeV values disallowed \\
\hline
0.08 & 0.4 & 6 & Figure \ref{exclusion_1}, except $\Delta m=10^{-14}$ GeV values disallowed \\
\hline
0.08 & 0.4 & 10 & Figure \ref{exclusion_2}, $\Delta m=10^{-14}$ GeV values disallowed \\
\hline
\end{tabular}
\end{center}
\caption{The description of other exclusion regions for different parameters when compared to Fig.\ \ref{exclusion_1} or \ref{exclusion_2}.}
\label{other_plots}
\end{table*}

In conclusion, sneutrino-antisneutrino oscillation is a very important tool to look for lepton number violation at the Tevatron. Even though, after the $\beta\gamma$ cut, the Lorentz factor $\gamma$ is not that high (and, even without the cut, the $\gamma$ is not that high at the Tevatron compared to the LHC, for example), the effects of the general oscillation probability formula need to be considered in order not to overestimate the oscillation probability in the very small sneutrino width scenario. This small a width ($\sim 10^{-14}$ GeV or so) can be realized in many different supersymmetric scenarios. A very interesting signal at the Tevatron could be two same-sign heavily ionized charged tracks and a soft pion, which can probe a mass splitting all the way down to $\sim 10^{-14}$ GeV with an integrated luminosity of 10 ${\rm fb}^{-1}$ for $L=0.28$ m, as can be seen from Table \ref{tab_cross_sec}. Various exclusion plots (Figs.\ \ref{exclusion_1}--\ref{exclusion_4} are shown) demonstrate whether the signal significance is $\geq 5$ for different $\Delta m$'s and $m_{\tilde{\nu}_\tau}$'s for different values of $m_{\tilde{\nu}_\tau}-m_{\tilde{\tau}_1}$. These plots show that there is an interesting $(\Delta m,m_{\tilde{\nu}_\tau})$ space available for the Tevatron collider.

\section*{Acknowledgements}

We thank T. Aaltonen for discussions. This work is supported in part by the Academy of Finland (Project No.\ 137960). D.K.G. acknowledges partial support from the Department of Science and Technology, India, under Grant No. SR/S2/HEP-12/2006. D.K.G. also thanks the ICTP High Energy Group for the hospitality. T.H. thanks the V\"ais\"al\"a Foundation for support.

\end{document}